# A Unified Probabilistic Framework for Spatiotemporal Passenger Crowdedness Inference within Urban Rail Transit Network

Min Jiang, Andi Wang, *Member, IEEE*, Ziyue Li, *Member, IEEE* and Fugee Tsung, *Member, IEEE*

*Abstract*— This paper proposes the Spatio-Temporal Crowdedness Inference Model (STCIM), a framework to infer the passenger distribution inside the whole urban rail transit (URT) system in real-time. Our model is practical since the model is designed in a probabilistic manner and only based on the entry and exit timestamps information collected by the automatic fare collection (AFC) system. Firstly, the entire URT system is decomposed into several components of stations and segments. By decomposing a passenger's travel actions into entering, traveling, transferring, and exiting, we build a statistical model to estimate the passengers' lingering time within each component and the passengers' destination based on historical AFC data. Then, the passengers' spatial distribution is predicted in real-time based on each passenger's elapsed travel time and their entry station. The effectiveness of the scheme is validated with a real dataset from a real URT system.

## I. Introduction

The urban rail transit (URT) network is a critical infrastructure to facilitate the daily commute of a large portion of residents. With its increasing expansion in many cities over the world, the crowdedness has been a major concern for URT networks which leads to delays and safety threats. As an important infrastructure of intelligent transport systems (ITS) [1], Understanding and monitoring the level of crowdedness of the URT are important for improving both efficiency and safety of URT networks. In recent decades, the improvement of data acquisition and management technologies has brought opportunities for crowdedness inference. With the wide application of automatic fare collection (AFC) systems, passengers' entry and exit information can be recorded once they swipe the smart cards at the gates. Therefore, it contains rich information about the passengers' travel patterns.

This paper aims to use AFC data for the crowdedness inference **within** the entire URT network. Specifically, we aim to infer the crowdedness, and the distribution of the passengers inside the URT system based on both the historical and the real-time AFC records. There are multiple challenges involved. First, only the information about entering and exiting the URT network is recorded by the AFC system, which means that the exact locations of the passengers are unknown. Second, we do not know the destination of the passengers who remain in the URT network from the real-time data. Besides, it is challenging to process millions of records generated every day. If real-time crowdedness estimation cannot be performed promptly, early warning of the crowding status cannot be achieved.

To solve the above challenges, we proposed a unified and probabilistic SpatioTemporal Crowdedness Inference Model (STCIM) that evaluates the passengers' spatial distribution across the whole URT network in a real-time manner. We apply the individual passenger travel action decomposition to the crowdedness inference: the URT network is decomposed into multiple types of components, namely (1) the stations, (2) transfer points, and (3) railway segments between stations. Each one-way trip is then divided into multiple actions of passengers, including entering, onboarding (moving), transferring, and exiting, and each action is associated with a component in the network and thereby contributes to the crowdedness of this location. To the best of our knowledge, we are the first model that could formulate the passengers' action time, location distribution, destination distribution, and crowdedness in station and segment components in a unified manner. Probabilistic models are carefully designed and constructed to prescribe each of these three action times, which enable the inference of the location of the passenger using the entry-exit time information. Then, the parameters in this model are estimated by the method of moment using historical data. With the estimated parameters, the probability of a passenger's location in each component is predicted with Monte Carlo method, and the number of passengers, i.e., the crowdedness within each component, is predicted.

The contributions of this paper are summarized as follows:

- We propose a unified model, i.e., STCIM, to infer the passenger crowdedness that is spatially distributed across the whole network in a real-time manner, and at the same time, it infers passenger's travel time, real-time location distribution, and destination distribution.

- We conduct abundant experiment in a real case study, demonstrating the model's ability to infer the crowdedness in station level, railway segment level, and the whole system level with visualization.

The remaining part of the paper is organized as follows. In Section 2, we review previous literature on AFC data and crowdedness analysis. Section 3 gives the details about the integrated analytical method that evaluates the crowdedness in URT networks. In Section 4, we demonstrate the proposed STCIM with a real URT system. Section 5 concludes our work and discusses potential research perspectives.

M. Jiang is with the University of Hong Kong, Hong Kong (e-mail: min.jiang@hku.edu). A. Wang is with Arizona State University, Tempe, AZ, USA (e-mail: andi.wang@asu.edu). Z. Li is with the University of Cologne, Germany (e-mail: zlibn@wiso.uni-koeln.de). F. Tsung is with Hong Kong University of Science and Technology (phone: (+852)-23587097; e-mail: season@ust.hk).

The authors will like to thank the Hong Kong MTR for providing the technical support.

## II. LITERATURE REVIEW

In the Transit Capacity and Quality of Service Manual [2], three indices are defined to evaluate the crowdedness: passenger flow, density, and speed. In our study, among others, density is used as the criteria to evaluate passengers' crowdedness, which relates to the number of passengers within a limited space. As our analysis considers the fixed operation areas within a URT network, the crowdedness can be defined as the number of passengers who are currently staying inside the network at the moment.

To evaluate the crowdedness in the area, some researchers [3, 4] use video information from Closed Circuit Television (CCTV) to count the passengers. However, privacy concerns may not support the installation of video cameras at every place in the URT network, and the video cameras will not cover the entire space, which only offers a partial estimation with potential underestimation. In addition, this CCTV-based method cannot be applied directly to predict future crowdedness levels.

Another way of estimating the number of passengers is to use tracking devices like cell phones with GPS and Wi-Fi modules [5]. However, this method also requires users' permission due to privacy concerns [6]. As a result, the system-wide information is unavailable for a large portion of passengers, and the crowdedness estimation will be affected.

Crowdedness inference in URT networks has been studied by operators of transit authorities. However, there is no existing approach that produces a holistic description of passengers' spatial and temporal distribution inside the URT system: most studies only focus on a local component of the URT network, such as carriage, platform, or railway segment. For example, Ceapa, et al. [7] conducted clustering analysis on stations in the Tube (the URT network in London) based on their congestion patterns. Caspari, et al. [8] proposed a crowdedness analysis method for a platform based on the estimation of travelers' entry directions and the information from the automated vehicle location system. This approach was validated in New York City Transit, and the estimations were very close to the evaluation from field surveys. A carriage-level study was conducted by Berkovich, et al. [9]. However, as multiple components in the URT network are interdependent with each other, and the passengers pass along their individual routes, modeling the crowdedness within a single network component is usually inadequate for system-wide analysis. To the best of our knowledge, there are only a few works trying to infer the crowdedness inside the public transport systems, such as buses [10]. The only system-wide study of crowdedness is Stasko, et al. [11], where passenger movement between different URT components is modeled by a spatial-temporal network. However, they do not take the uncertainty of walking time and waiting time into consideration; instead, they strictly assume the lingering time of a passenger within a component being constant.

This paper contributes to the existing literature in the following aspects. (1) The entry and exit information of all stations is considered when modeling the crowdedness of the entire URT network. (2) The uncertainty of traveling time and waiting time is taken into consideration. (3) A visualization system-wide crowdedness level within the entire

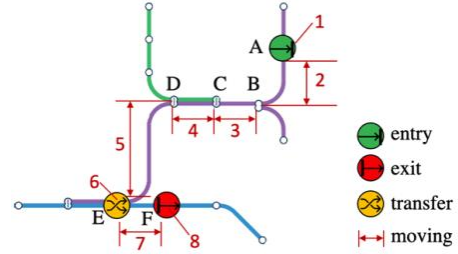

Figure 1  A trip from Station A to Station F in a URT system

URT network.

Last but not the least, we would like to highlight the difference between our work and demand prediction [12, 13]. Demand prediction in the URT context means to predict how many passengers entering one specific station, whereas ours model infers how many passengers are currently staying at this station. More discussions are in Sec. IV.B.1) and 2). The recent individual analysis could offer a potential solution, which directly deal with individual passenger travel data, such as OD travel pattern [14, 15] or trajectory [16]. However, the current individual analysis only tried to understand the passenger mobility pattern and behavior, not the crowdedness inference.

## III. METHODOLOGY

In this section, we propose an integrated analytical method, *i.e.*, STCIM, for the crowdedness inference within the URT network.

We first define the following *components* for a URT network to model the passengers' movement inside it. In a URT network $G = (V, E)$ with $J$ service lines, the set of *stations* is represented by $V = \{1, ..., I\}$, and each edge $(s_1, s_2) \in E$ is a *railway segment* linking two stations of $s_1$ and $s_2$. Service *line* $j$ is represented by a subgraph $G_j = (V_j, E_j)$, where $V_j = \{s_{j,1}, ..., s_{j,I_j}\} \subset V$ is the set of all stations on line $j$, and it contains $I_j - 1$ railway segments $E_j = \{(s_{j,1}, s_{j,2}), ..., (s_{j,I_j-1}, s_{j,I_j})\}$. We assume that each segment belongs to only one line, while a *transfer station* $\tilde{s} \in S$ is a common station of two or more lines where the passengers may change from one line to another and the set of transfer stations $\tilde{S}$.

With the above definition, every trip from one station $s$ of line $j$ to another station $s'$ of line $j'$ can be decomposed as a sequence of *actions*, each corresponding to one of the following categories: (1) entering a station $s$ of line $j$; (2) moving along a railway segment $(s_{j,k}, s_{j,k+1}) \in E_j$; (3) transferring from one line $\tilde{j}$ to another line $\tilde{j}'$ at station $\tilde{s} \in V_{\tilde{j}} \cap V_{\tilde{j}'}$; (4) leaving the URT network from station $s'$ of line $j'$.

For example, the entire trip from station A to station F, as shown in Figure 1, is composed of the following eight actions: 1. Enter station A; 2. Move from A to B; 3. Move from B to C; 4. Move from C to D; 5. Move from D to E; 6. Transfer from the purple line to the blue line at station E; 7. Move from station E to station F; and 8. Leave the URT network from station F. For most URT networks, there is a unique optimal transit route for a trip from an entry station to an exit station, with the minimum travel time between them and corresponding to a specific sequence of actions.

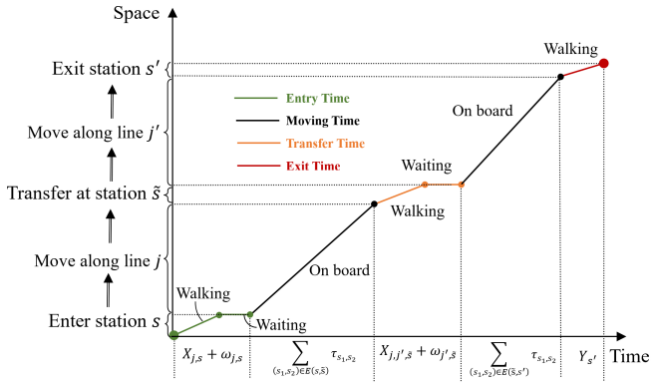

Figure 2 Spatial and temporal movement of passengers' actions along the route

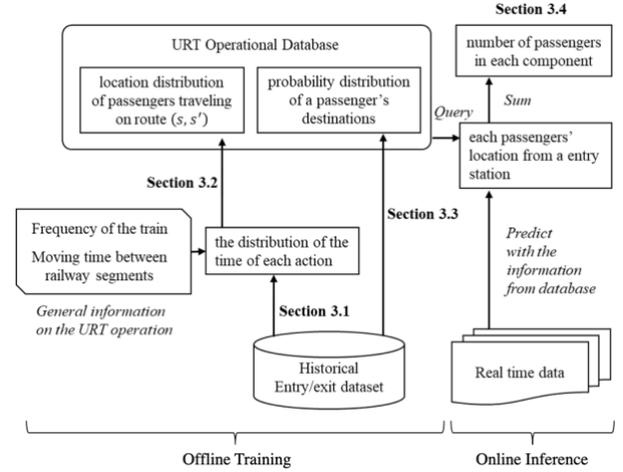

Figure 3 Flowchart of implementing the system-wide estimation of crowdedness

Let the actions along the route $(s, s')$ be $a_1^{(s,s')}, a_2^{(s,s')}, \ldots, a_{m_{s,s'}}^{(s,s')}$, where $m_{s,s'}$ is the number of actions. In general, the first action of the route $a_1^{(s,s')}$ is entering station $s$, which is in category (i). The last action $a_{m_{s,s'}}^{(s,s')}$ is leaving the URT network from station $s'$, which is in category (4). Other actions in the middle of the route are in categories (1) and (3).

Specifically, the set of actions of category (2), moving along a railway segment is represented as $E(s, s') \subset E$. Similarly, all transfers along the route from station $s$ to $s'$ is represented as set $\tilde{S}(s, s')$, where each element in it is represented by $(\tilde{s}, \tilde{j}, \tilde{j}')$, denoting the transfer from line $\tilde{j}$ to line $\tilde{j}'$ at station $\tilde{s}$.

To evaluate crowdedness within the entire URT network, predicting the location of passengers is critical, which in turn relates to the duration of time that passengers stay in every network component. In our STCIM model, we propose to estimate the duration of passengers' actions using historical data. The uncertainty of a passenger's action time in the URT network can be described as follows (illustrated in Figure 2).

- **Entry Time**: The time that a passenger enters station $s$ to take a train along line $j$ can be decomposed into two intervals: the *pure entry time* from a passenger taps in at the entrance station to the time when this passenger arrives at the platform of line $j$, and the *waiting time*, which is the duration that the passenger stays on the platform and wait for the next train. The pure entry time at station $s$ of line $j$ is represented by a random variable $X_{j,s}$. As the randomness of interarrival time between two trains is small, the waiting time $\omega_{j,s}$ can be characterized by a uniform distribution $U(0, \delta_j)$. Specifically, we assume that the duration that a passenger walks to the platform is only dependent on the passenger's walking habit and the distance, and not affected by the train's schedule, which indicates that $\omega_{j,s}$ is independent with $X_{j,s}$, where $\delta_j$ is the inter-arrival time of the train at line $j$ at time $t$. In total, entry time can be represented as $X_{j,s} + \omega_{j,s}$.

- **Moving Time**: The time that the passenger stays on the train while moving along a segment of $(s_1, s_2) \in E(s, s')$ is a fixed value, denoted as $\tau_{s_1, s_2}$. It includes the time that the train stops at a station. As the uncertainty of each $\tau_{s_1, s_2}$ is low, it is assumed to be a time-independent constant [17] that can be obtained from the URT operator.

- **Transferring Time**: The time for transferring from line $\tilde{j}$ to $\tilde{j}'$ at station $\tilde{s}$ contains the walking time from one platform to another and the waiting time at the second platform. For each time of transfer at a transfer station $m$, the total transfer time is $X_{\tilde{j}, \tilde{j}', \tilde{s}} + \omega_{\tilde{j}', \tilde{s}}$, where $X_{\tilde{j}, \tilde{j}', \tilde{s}}$ is walking time from the arrival platform of line $\tilde{j}$ to the departure platform of line $\tilde{j}'$ in station $\tilde{s}$, and the waiting time $\omega_{\tilde{j}', \tilde{s}} \sim Unif(0, \delta_{\tilde{j}'})$. Thus, the total transferring time is $X_{\tilde{j}, \tilde{j}', \tilde{s}} + \omega_{\tilde{j}', \tilde{s}}$.

- **Exit time:** The duration that a passenger walks from the platform to the gate in the exit station. For this reason, the distribution of the exit time is $X_{j', s'}$, which is the same as the pure entry time at station $s'$ of line $j'$, as the entry and exit gates are usually very close in the station.

In above formulations, we need to specify the distribution of pure entry time of every station and the distribution of the transfer time, i.e., the distribution of $X_{j,s}$'s for all station $s$ on line $j$, and the distribution of $X_{\tilde{j}, \tilde{j}', \tilde{s}}$'s for all transfers from line $\tilde{j}$ to $\tilde{j}'$ at station $\tilde{s}$. We characterize these random variables with Gamma distributions, as it is a unimodal distribution suitable for nonnegative, continuous random variables, and provides the flexibility on modeling both the mean and the variance of the passengers' walking time [18]. Note that our STCIM allows for using other distributions, like the truncated normal distribution [19], to model walking time, as long as the distribution family can be characterized by its mean and variance. Besides, it is worth mentioning that the assumptions are valid only if the platform and the trains are not overcrowded, so that (1) the travel speed is not related to the number of passengers in the component, (2) all passengers are guaranteed to board the first train that arrives at the platform.

The STCIM model is illustrated in Figure 3. Firstly, we obtain the distribution of passengers' action time in Section

III.A. The information on the action time will be used to obtain the probability distribution of the locations of a passenger who travels from station $s$ to station $s'$ for time $\Delta t$. This is achieved through a Monte Carlo method, as described in Section III.B. We also obtain the passengers' destination distribution starting from any station $s$, described in Section III.C. The information obtained from Section III.B and C are stored in an operational database. During the operation of URT, the STCIM model predicts each passenger's location and the total number of passengers in each URT network component based on the information from the real-time AFC data, as introduced in Section III.D.

## A. Estimating the passengers' action time

Recall that the passengers' actions include waiting on platforms, moving along a railway segment on the train, entering stations, exiting stations, and transferring trains. From the general information about the URT operation, we can directly obtain the values of the trains' interarrival time $\delta_j$'s and the moving time $\tau_{s_1,s_2}$ along a railway segment linking station $s_1$ and $s_2$. To obtain the distribution of passengers' walking time, we introduce how to estimate the parameter of the walking time distribution $X_{j,s}$ and $X_{j,j',\tilde{s}}$'s from the historical dataset recording passengers' entry and exits time stamps.

Let $\mathbf{X} \in \mathbb{R}^n$ be the vector containing all $X_{j,s}$'s and $X_{j,j',\tilde{s}}$'s defined on the URT network and let the $l$-th element $[\mathbf{X}]_l := X_l \sim \text{Gamma}\left(\theta_l^{[1]}, \theta_l^{[2]}\right)$. We estimate the parameter $\theta_l^{[1]}, \theta_l^{[2]}$ from the historical dataset containing all the transit records in given period using the method of moment. Let there be $N_{s,s'}$ trips between an entry station $s$ (on line $j$) and an exit station $s'$ (on line $j'$), and the entry and exit time stamps of trip $k$ are denoted as $t_{s,s',k}^{\text{in}}$ and $t_{s,s',k}^{\text{out}}$, $k = 1,2,\ldots,N_{s,s'}$. From the data, the travel times can be calculated as $T_{s,s'}^{(k)} = t_{s,s',k}^{\text{out}} - t_{s,s',k}^{\text{in}}$, and we obtain their sample mean and sample variance as follows.

$$\hat{\mu}_{s,s'} = \frac{1}{N_{s,s'}} \sum_{k=1}^{N_{s,s'}} T_{s,s'}^{(k)}; \hat{\sigma}_{s,s'}^2 = \frac{1}{N_{s,s'}} \sum_{k=1}^{N_{s,s'}} \left(T_{s,s'}^{(k)} - \hat{\mu}_{s,s'}\right)^2 \quad (1)$$

Now, we calculate the population mean and variance of each $T_{s,s'}^{(k)}$ based on the passengers' movement model. Based on the assumptions mentioned in Section 3.1, $T_{s,s'}$, the total travel time from $s$ to $s'$, is characterized by the summation of entry time, moving time on the train, transfer time, and exit time. Thus, the total travel time can be given as follows:

$$T_{s,s'} = (X_{j,s} + \omega_{j,s}) + \sum_{(s_1,s_2) \in E(s,s')} \tau_{s_1,s_2}$$
$$+ \sum_{(\tilde{s},\tilde{j},j') \in \tilde{S}(s,s')} (X_{\tilde{j},j',\tilde{s}} + \omega_{\tilde{j}',\tilde{s}}) + X_{j',s'}. \quad (2)$$

Its expected value is as follows:

$$\mathbb{E}[T_{s,s'}] = \mathbb{E}[X_{j,s}] + \frac{\delta_j}{2} + \sum_{(s_1,s_2) \in E(s,s')} \tau_{s_1,s_2}$$
$$+ \sum_{(\tilde{s},\tilde{j},j') \in \tilde{S}(s,s')} \left(\mathbb{E}[X_{\tilde{j},j',\tilde{s}}] + \frac{\delta_{\tilde{j}'}}{2}\right) + \mathbb{E}[X_{j',s'}] \quad (3)$$

Similarly, the variance of travel time $\text{var}[T_{ij}]$ can be expressed as:

$$\text{var}[T_{s,s'}] = \text{var}[X_{j,s}] + \frac{\delta_j^2}{12}$$
$$+ \sum_{(\tilde{s},\tilde{j},j') \in \tilde{S}(s,s')} \left(\text{var}[X_{\tilde{j},j',\tilde{s}}] + \frac{\delta_{\tilde{j}'}^2}{12}\right) + \text{var}[X_{j',s'}] \quad (4)$$

Note that the right-hand side of either Eq. (3) or Eq. (4) is a linear function over $\boldsymbol{\mu}_{\mathbf{x}} = (\mathbb{E}[X_1], \ldots, \mathbb{E}[X_n])$ or $\mathbf{s}_{\mathbf{x}} = (\text{var}[X_1], \ldots, \text{var}[X_n])$ plus a constant, and thus we can represent them as $c_{s,s'}^\mu + \mathbf{r}_{s,s'}^\top \boldsymbol{\mu}_{\mathbf{x}}$ and $c_{s,s'}^s + \mathbf{r}_{s,s'}^\top \mathbf{s}_{\mathbf{x}}$.

To estimate the parameters $\boldsymbol{\theta}^{[1]}, \boldsymbol{\theta}^{[2]}$, we match the moments of population in Eq. (3) and Eq. (4) with the sample's moments in (1) for all $s, s'$. In total, there are $I(I-1) = O(I^2)$ pairs of entry-exit stations. However, the number of parameters is $n$, which is typically much smaller. To solve this challenge, we obtain the estimation of $\boldsymbol{\mu}_{\mathbf{x}}$ and $\mathbf{s}_{\mathbf{x}}$ by solving the following weighted least square problems:

$$\min_{\boldsymbol{\mu}_{\mathbf{x}}} \sum_{s,s'} (\hat{\sigma}_{s,s'}^2 / N_{s,s'})^{-1} \left(\hat{\mu}_{s,s'} - c_{s,s'}^\mu - \mathbf{r}_{s,s'}^\top \boldsymbol{\mu}_{\mathbf{x}}\right)^2,$$
$$\min_{\mathbf{s}_{\mathbf{x}}} \sum_{s,s'} (1/N_{s,s'})^{-1} \left(\hat{\sigma}_{s,s'}^2 - c_{s,s'}^s - \mathbf{r}_{s,s'}^\top \mathbf{s}_{\mathbf{x}}\right)^2. \quad (5)$$

The problems aim to match $\hat{\mu}_{s,s'}$ with $c_{s,s'}^\mu + \mathbf{r}_{s,s'}^\top \boldsymbol{\mu}_{\mathbf{x}}$ and match $\hat{\sigma}_{s,s'}^2$ with $\mathbf{r}_{s,s'}^\top \mathbf{s}_{\mathbf{x}}$ for all routes $(s, s')$. Note that the weights $(\hat{\sigma}_{s,s'}^2 / N_{s,s'})^{-1}$ and $(1/N_{s,s'})^{-1}$ are applied, because the sampling distributions of $\hat{\mu}_{s,s'}$ and $\hat{\sigma}_{s,s'}^2$ are with the variance [20]

$$\text{var}(\hat{\mu}_{s,s'}) = \frac{1}{N_{ij}} \text{var}[T_{s,s'}] \sim \frac{1}{N_{ij}} S_{ij},$$
$$\text{var}(\hat{\sigma}_{s,s'}^2) = \frac{1}{N_{ij}} \left(\mathbb{E}[T_{s,s'} - \mu_{s,s'}]^4 - \frac{n_{ij} - 3}{n_{ij} - 1} (\text{var}[T_{s,s'}])^2\right) \propto \frac{1}{N_{ij}} \quad (6)$$

respectively, and the weight of each route $(s, s')$ is inversely proportional to $\text{var}(\hat{\mu}_{s,s'})$ and $\text{var}(\hat{\sigma}_{s,s'}^2)$ respectively. After $\hat{\boldsymbol{\mu}}_{\mathbf{x}}$ and $\hat{\mathbf{s}}_{\mathbf{x}}$ are obtained, we will solve $\theta_l^{[1]}$ and $\theta_l^{[2]}$ from Eq. (7) given to the property of the Gamma distribution.

$$[\hat{\boldsymbol{\mu}}_{\mathbf{x}}]_l = \theta_l^{[1]} \theta_l^{[2]}; [\hat{\mathbf{s}}_{\mathbf{x}}]_l = \theta_l^{[1]} \theta_l^{[2]^2} \quad (7)$$

## B. Passengers' location distribution along a route

To predict the passengers' location distribution within the network, we first predict the location of the passenger along a specific route $(s, s')$ who enters for time $\Delta t$ using Monte Carlo simulation. Specifically, we calculate $P(a|s, s', \Delta t)$, the probability that a passenger is performing action $a$ given known route $(s, s')$ as this passenger has entered the station $s$ for time $\Delta t$ and remains in the URT network. We use Algorithm 1 to calculate $P(a|s, s', \Delta t)$, for a series of time $\Delta t = r\delta$ for $r = 1,2,\ldots,R$, where $\delta$ is a small time unit. In the beginning, we initiate zero matrix $\mathbf{A}$ of size $(m_{s,s'} + 1) \times R$. In lines 2-6 of this algorithm, we generate $I_{\max}$ trips along route $(s, s')$, and use the $(c, r)$-element $A_{c,r}$ to count the number of trips in which the passenger is performing the action $a_c^{(s,s')}$ at time $r\delta$ when $c \leq m_{s,s'}$, and $A_{m_{s,s'}+1,r}$ is reserved for counting the cases in which the passenger leaves the URT network at time $r\delta$. The probability that the passenger is performing $a_c^{(s,s')}$ at time $r\delta$, given that this passenger has not exited the URT network is thus the ratio of

**Algorithm 1**: Calculate $P(a|s, s', \Delta t)$

---
1 Initiate number of iterations $I_{\max}$ and zero matrix $\mathbf{A} \in N^{(m_{s,s'}+1) \times R}$
2 For $i = 1, \dots, I_{\max}$:
3 | For $c = 1, 2, \dots, m_{s,s'}$:
4 | | Generate $\phi_c^i$ according to the distribution of action time for $a_c^{(s,s')}$.
5 | | $A_{c,r} \leftarrow A_{c,r} + 1$ for all $r$ such that $\sum_{l=1}^{c-1} \phi_l^i < \delta r \leq \sum_{l=1}^{c} \phi_l^i$
6 | $A_{(m_{s,s'}+1),r} \leftarrow A_{(m_{s,s'}+1),r} + 1$ for all $j$ such that $\delta r > \sum_{l=1}^{c} \phi_l^i$
7 For $r = 1, \dots, R$ and $c = 1, 2, \dots, m_{s,s'}$:
8 | $P\left(a_c^{(s,s')} \middle| s, s', r\delta\right) \leftarrow \frac{A_{c,r}}{A_{1,r} + \dots + A_{m_{s,s'},r}}$;

---

the total number of trips where the passenger is performing action $a_c^{(s,s')}$ at time $r\delta$, and the total number of trips where the passenger remains in the URT network at time $r\delta$, calculated in line 7-8.

*C. Estimating the destination distribution*

In crowdedness inference, the number of passengers in each component is only dependent on the passengers who have entered but have not exited the URT network. However, the real-time data only include the entry time and entry station for these passengers. To predict their locations in the URT network, it is thus essential to predict their destinations.

Consider a passenger who enters the URT network from station $s$ at time $t^{\text{in}}$ and stays in it for time over time $\Delta t$. Then, $P(s'|s, t^{\text{in}}, \Delta t)$ is the probability that this passenger aims to travel to station $s'$. We estimate this probability through the frequency of such trips in the historical data set. Specifically, we first find $N_{s,s',t^{\text{in}},\Delta t}$, the number of records with entering station $s$, exit station $s'$, entering time around $t^{\text{in}}$, and travel time over $\Delta t$ from the historical AFC dataset. Then we estimate $P(s'|s, t^{\text{in}}, \Delta t)$ with $\frac{N_{s,s',t^{\text{in}},\Delta t}}{\sum_{s'} N_{s,s',t^{\text{in}},\Delta t}}$.

*D. Crowdedness Evaluation*

As the values of $P(a|s, s', \Delta t)$ for $s, s' \in S$, $a$, and $\Delta t = \delta, 2\delta, \dots, R\delta$ are obtained in Section 3.2 and $P(s'|s, t^{\text{in}}, \Delta t)$ is obtained in Section 3.3, we can further calculate $P(a|s, t^{\text{in}}, \Delta t)$, the probability that a passenger who entered station $s$ at time $t^{\text{in}}$ for duration $\Delta t$ is performing action $a$.

$$P(a|s, t^{\text{in}}, \Delta t) = \sum_{s' \in S} P(a|s, s', \Delta t) P(s'|s, t^{\text{in}}, \Delta t). \quad (8)$$

This equation allows us to predict the number of passengers performing each action in the network in real-time. Let $t^{\text{now}}$ be the current time, the number of passengers who entered station $s$ and remained in the URT network be $N_s(t^{\text{now}})$, and let the entry time of these passengers be $t_{s,i}^{\text{in}}$ for $i = 1, \dots, N_s(t^{\text{now}})$. According to Eq. (), the probability that a passenger $i$ entered station $s$ is performing action $a$ can be computed as $P(a|s, t_{s,i}^{\text{in}}, t^{\text{now}} - t_{s,i}^{\text{in}})$. Then, the expected value of $C_{a,t^{\text{now}}}$, the total number of passengers performing action $a$ at moment $t^{\text{now}}$ in the entire URT network, is thereby expressed as:

$$\mathbb{E}[C_{a,t^{\text{now}}}] = \sum_{s \in S} \sum_{i=1}^{N_s(t^{\text{now}})} P(a|s, t_{s,i}^{\text{in}}, t^{\text{now}} - t_{s,i}^{\text{in}}). \quad (9)$$

To evaluate the crowdedness within a specific URT network component, we can calculate the sum of all $\mathbb{E}[C_{a,t^{\text{now}}}]$ whose actions $a$ are associated with this component. As a result, we obtained a real-time evaluation of the crowdedness in every component of the URT network, and this evaluation can be used for visualizing the crowdedness inside the entire URT network, as will be seen in Section 4.2.

*E. Implementation details*

In our STCIM model, estimating the action time parameters in Section 3.1 assists calculating the probability distribution of the current action $P(a|s, s', \Delta t)$ along a route in Section 3.2. The probability distribution of the destination $P(s'|s, \Delta t)$ is independently calculated in Section 3.3. All the computations in Section 3.1, 3.2 and 3.3 rely on the historical dataset, while the major computational burden comes from Section 3.2. The computation in these subsections generates the probability $P(a|s, s', \Delta t)$ and $P(s'|s, t^{\text{in}}, \Delta t)$ for all routes $(s, s')$, incoming time $t^{\text{in}}$, and current travel time $\Delta t$. To alleviate the computational burden in real-time crowdedness prediction, we propose to calculate these values offline from the historical data and store these values in a URT operational database. The operators may perform the algorithms in Section 3.1, 3.2 and 3.3 to update the database on a regular basis (for example, once a month).

In the real-time monitoring process, the crowdedness within the URT network is predicted with the method in Section 3.4. At each time $t^{\text{now}}$, the probability distribution of the actions for each passenger inside the URT network can be predicted with Eq. (8), by querying $P(a|s, s', \Delta t)$ and $P(s'|s, t^{\text{in}}, \Delta t)$ from the database. Then, the crowdedness can be evaluated via Eq. (9), and the time complexity is $O(I \sum_{s \in S} N_s(t^{\text{now}}))$, where $\sum_{s \in S} N_s(t^{\text{now}})$ is the total number of passengers in the URT network and $I$ is the total number of stations. This computational complexity is linear to the total number of passengers, which enable rapid evaluation of the crowdedness.

## IV. APPLICATION

We apply the proposed methodology to a real URT system, to demonstrate the proposed STCIM in evaluating and visualizing the crowdedness. In the entire URT network, there are about five million passenger rides on average in 2020. Therefore, the case study is representative enough to examine the efficiency and effectiveness of our proposed method. We obtained seven days' transit data. Due to privacy concerns, the data we obtained are generated from private simulation platform used by this URT company, instead of the real AFC records. Although the passenger flow of each route across the day is very close to the reality, this simulation platform generates trips from one station $s$ to another station $s'$ with the same travel time, and thereby in our simulation we perturb the walking time with a variance of 5s² for all stations. It is worth mentioning that despite the trips of these seven days are different, the travelling time of passengers are not significantly different, so our approach is still applicable. To demonstrate our methodology, we modify this dataset by incorporating the randomness of travel time in the URT network based on our distribution assumptions in Section 3. In our modified dataset, the average travel time of all trips along a route is equal to the constant travel time in the original dataset approximately. Due to this limitation of the simulated data, we do not aim at validating the distribution assumptions in this article, and we recognize it as a potential future research topic. Instead, we focus on demonstrating the capability of crowdedness evaluation and visualization in this article.

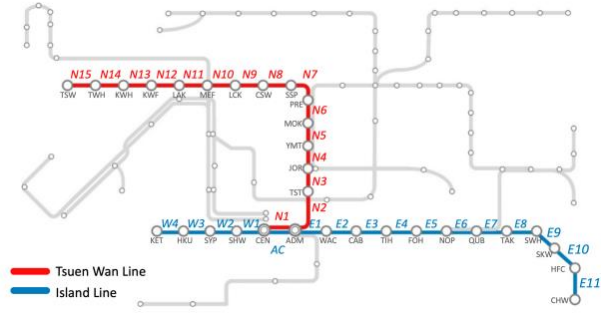

Figure 4 Blue and red line of the URT network, with stations (grey) and rail segments (blue, red)

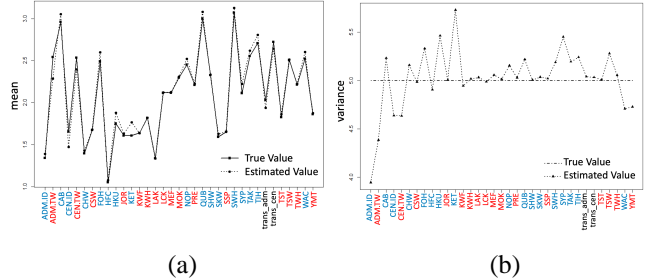

(a)                 (b)

Figure 5 True and estimated values of (a) the mean and (b) the variance of walking time (in seconds).

With the modified dataset, we use the first six days' data as the historical data and use the last day's data for real-time crowdedness estimation. The network consisting of $J = 2$ lines is shown in Figure 4: they contain $I = 31$ stations (two of them are transfer stations), which are among the busiest lines of the entire URT network we considered, covering more than one million transits per day and accounting for over 1/4 of transits within the entire URT network. The railway segments of these two lines are marked as "W1", ..., "W4", "E1", ..., "E11", "N1", ..., "N15", and "AC". Note that both red line and blue line have a railway segment between station CEN and ADM. In the URT network, the passengers coming from the north red line travel along segment N1 and transfer at CEN to go westward through blue line, and passengers coming from the west blue line transfer to red line at ADM if they are going north. This special design is incorporated in our definition of routes between stations.

*A. Travel time estimation*

In the URT network that we consider, there are 31 stations. Therefore, the number of routes is $31 \times (31 − 1) = 930$. From all routes, we calculate the mean and variance of the travel time, with stations in blue line in text color blue and stations in red line in text color red. Based on these results, we calculate the mean and variance of walking time by solving the optimization problems in Section III.A. Note that due to the special layout of the transfer stations ADM or CEN, exiting from the platform of the red line or blue line will incur different exit time distributions. Therefore, walking from two platforms in each station to the gate are regarded as two actions, specified by ADM.ID in blue line and ADM.TW in red line (or CEN.ID in blue line and CEN.TW in red line). Also, the transfer actions between two lines at either Admiralty or Central station are specified as "trans_adm" and "trans_cen" individually (in text color black). In Figure 5 (a) and (b), we show the estimated mean and variance of the walking time, i.e., $\hat{\mu}_x$ and $\hat{s}_x$. The estimated values of $\hat{\mu}_x$ in Figure 5 (a) is compared with that of the true mean specified in our simulation platform, and we found the estimation quite good for most stations. The mean squared error of estimating the mean walking time $\hat{\mu}_x$ across all stations is $7.12 \times 10^{-3}$ and the mean squared error of estimating the variance of walking time $\hat{s}_x$ is 0.096. After the mean and variance of the staying time is obtained, the parameters of the Gamma distributions are estimated. We obtained the staying time of passengers for every railway segment and the travel time for every railway segment from the schedule of the URT network.

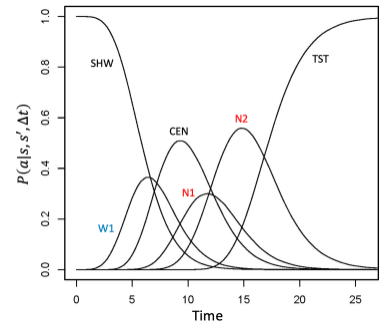

Figure 6 Probability distribution of a passenger's action at time $\Delta t$ (SHW-TST)

We then go through the procedure of Section III.B to find the probability distribution of passengers' action, and we illustrate the result for an example route from SHW to TST. This route includes six actions: entering SHW, moving along W1, transfer at CEN, moving along N1, moving along N2, and exiting from TST. Figure 6 illustrates the result of $P(a|s, s', \Delta t)$, where the y-axis denotes the probability value and the $x$-axis denotes $\Delta t$. When $\Delta t$ is close to 0, the passenger has a high probability of staying in SHW station, waiting for the train. Fifteen minutes after this person enters the network from SHW, the probability that the passenger is at SHW or W1 is quite small, and the most probable location of this passenger is in railway segment N2.

*B. Visualizing passengers' distribution within the URT network*

According to the proposed STCIM in Section 3, the information of destination distribution and the location distribution along individual routes is saved in database, to be used for evaluating the crowdedness in the URT system with the real time AFC data. We use the modified dataset of day 7 as the real-time data. At the time points from 8 a.m. to 8 p.m. with 10 min interval, we calculate the numbers of passengers in every component of the network with the method of Section III.D. As all data sets only provides the entry/exit timestamps of individual passengers, the ground truth of the crowdedness within the system is not available. Therefore, we illustrate the crowdedness prediction of various network components at different time, to validate that the crowdedness prediction reflects the reality of the system from multiple aspects.

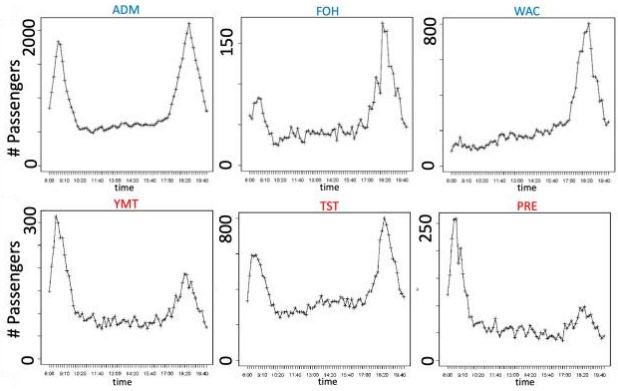

Figure 7 Crowdedness in Several Representative Stations

*1) Crowdedness in Each Station*

Based on the result, we first show the variation of crowdedness of six stations throughout one day. Figure 7 shows the total number of passengers who are staying in six representative stations: ADM, TST, PRE, YMT, FOH, and WAC. These stations can be divided into four groups based on the trend of crowdedness in a day. First, ADM and TST are bimodal with a similar level of morning peak and evening peak. Second, PRE and YMT have a morning peak and a much lower evening peak. Third, stations like FOH are the opposite and have a higher evening peak. Fourth, stations like WAC only has one evening peak in a day. There is no station without evening peak from our finding. We would like to emphasize the difference between the crowdedness in each station and the passenger inflow or outflow value in each station Li, et al. [12], [13]. Crowdedness measures how many passengers is currently staying in this station, whereas inflow or outflow value measures how many passengers are entering or exiting this station.

*2) Crowdedness in Each Railway Segment*

We evaluate the crowdedness of each segment of railway in different times within a day, which is essential for operation and scheduling. Figure 8 shows the number of passengers in each station and each railway segment along red line and blue line in the morning (8:30 am), at noon (11:30 am), and in the evening (6:30 pm). We find that the southern stations along the red line serve more passengers than the northern stations while the busiest segments of blue line are in the middle part. Comparing the morning peak with evening peak, we can see the patterns of red line are quite similar. However, the blue line is more crowded during the evening peak.

Here we also would like to highlight the difference between the crowdedness in each railway segment and the passenger flow between an origin-destination (OD) pair. Take the railway segment between stations A and B as an example, the crowdedness in the railway segment measures how many passengers are physically located in the railway segment between A and B, but they do not necessarily only travel from A to B: they could possibly travel to other stations but just pass by the A-B rail segment; they could also possibly travel in the opposite direction such as from B to A. In contrast, the passenger flow between OD pair measures strictly the number

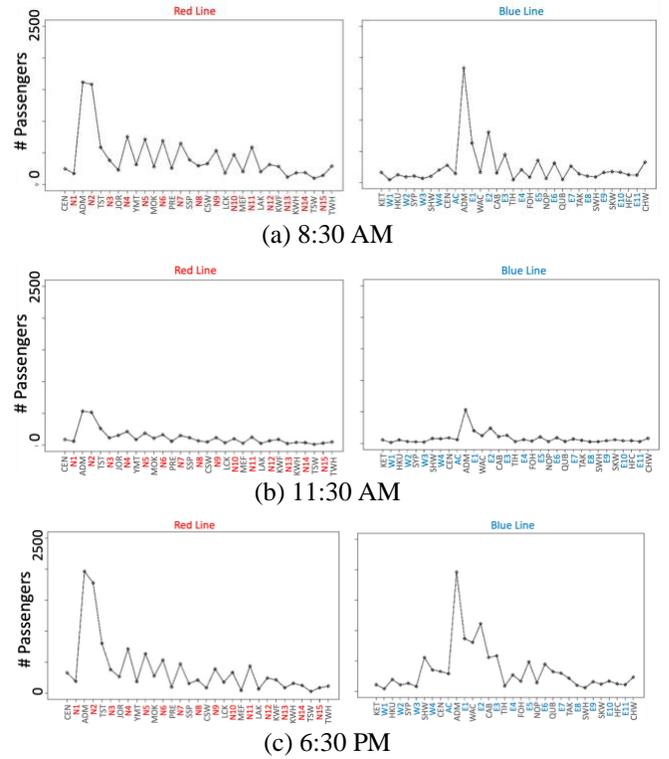

(a) 8:30 AM

(b) 11:30 AM

(c) 6:30 PM

Figure 8 Crowdedness along red line (left) and blue line (right) at different time, with stations in text color grey, and red line segments in red, and blue line segments in blue.

of passengers traveling from the origin (station A) to the destination (station B).

*3) Visualization for the System-wide Crowdedness Distribution*

Finally, we developed a system-wide visualization tool to show the crowdedness distribution of passengers inside the entire URT network, as illustrated in Figure 9. In this visualization tool, the color of each station and each line segment indicates the crowdedness of these components. The crowdedness of transfer stations is illustrated separately with the blocks beside the two transfer stations marked CEN/Trans and ADM/Trans, to shed some light on the cause of crowdedness of these stations. Figure 8 visualizes the crowdedness of the URT network at several times in a day from 8:30 to 20:00. This visualization provides rich information for the operator on the entire URT network. For example, we can see that the most crowded railway segment within a day is N2, linking TST to ADM. It is because these two stations connect two major downtown areas of the city.

In addition, this visualization clearly illustrates the temporal variation of the crowdedness within each component.

## V. CONCLUSION AND DISCUSSIONS

In this paper, we proposed an integrated analytical method, *i.e.*, SpatioTemporal Crowdedness Inference Model (STCIM) that uses the automatic fare collection (AFC) records obtained from both historical and online data for evaluating and visualizing the system-wide crowdedness within a URT network in real-time. The proposed STCIM integrates offline historical data and the real-time operational data from the AFC records for the purpose of system-wide crowdedness evaluation and visualization. The STCIM is based on estimating the passengers' locations inside the URT network. Specifically, it decomposes each trip between two stations into multiple actions of the passengers. Based on this model, we use historical data to extract useful information on the passengers' travel patterns.

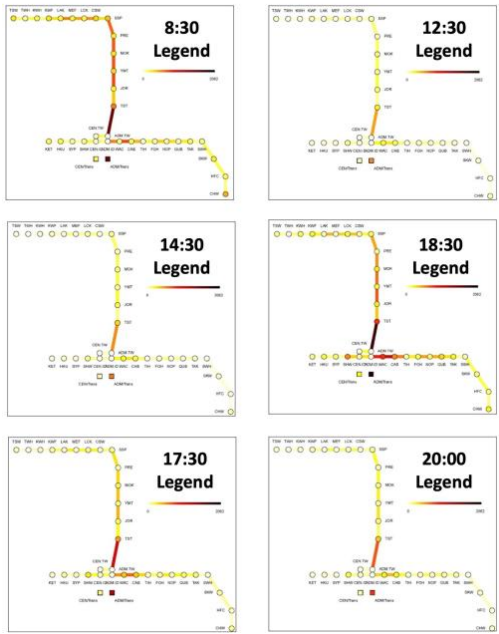

Figure 9 The crowdedness visualization across the entire URT network

Firstly, we estimate the time distribution of every action, which further provides information on the passengers' location inside the URT network along a route. Secondly, we obtain the destination distributions. The knowledge of the passengers from the historical data then helps to overcome the major deficiency of the real-time AFC data: the missing information of the passenger's destination and the current location. By combining historical data and real-time data, our STCIM model achieves efficient real-time crowdedness evaluation.

We demonstrate our STCIM with the AFC dataset of a URT network. With our method, we discovered how the crowdedness in the URT network varies throughout the day and across all stations. The information obtained from our STCIM model is critical for follow-up decision-making problems, including crowdedness monitoring, resource allocation, and scheduling.